\documentclass[11pt]{elsart}
\usepackage[dvips]{graphicx}

\begin{document}
\begin{frontmatter}

\title{Growing Scale-free Small-world Networks with Tunable Assortative Coefficient
}

\author{Qiang Guo\corauthref{jg}}
\corauth[jg]{Corresponding author. Tel. (+86)13050537943.}
\ead{liujg004@yahoo.com.cn}
\author{$^{a}$, Tao Zhou$^{b}$,}
\author{Jian-Guo Liu$^{c}$,}
\author{Wen-Jie Bai$^{d}$,}
\author{Bing-Hong Wang$^{b}$ and}
\author{Ming Zhao$^{b}$}

\address{$^{a}$ School of Science, DaLian Nationalities University,
Dalian 116600, P R China}
\address{$^{b}$ Department of
Modern Physics, University of Science and Technology of China, Hefei
Anhui, 230026, P R China}
\address{$^{c}$ Institute of System Engineering, Dalian University of Technology,
Dalian 116023, P R China}
\address{$^{d}$ Department of Chemistry, University of Science and Technology of China, Hefei
Anhui, 230026, P R China}


\begin{abstract} \textnormal{\small {In this paper, we propose a
simple rule that generates scale-free small-world networks with
tunable assortative coefficient. These networks are constructed by
two-stage adding process for each new node. The model can reproduce
scale-free degree distributions and small-world effect. The
simulation results are consistent with the theoretical predictions
approximately. Interestingly, we obtain the nontrivial clustering
coefficient $C$ and tunable degree assortativity $r$ by adjusting
the parameter: the preferential exponent $\beta$. The model can
unify the characterization of both assortative and disassortative
networks.}}

\end{abstract}
\date{}
\end{frontmatter}

\section{Introduction}
\medskip
In the past few years, no issues in the area of network researching
attract more scientists than the ones related to the real networks,
such as the Internet, the World-Wide Web, the social networks, the
scientific collaboration and so on
\cite{WS98,BA99,AB02,DM02,New,XFWang01}. Recent works on the complex
networks have been driven by the empirical properties of real-world
networks and the studies on network dynamics
\cite{E1,E2,E5,EE5,E6,E7,E8,E9,E10,D1,D2,D3,D4,D5,D6,D7,D8,D9,D10,D11,D12,D13,D14}.
Many empirical evidences indicate that the networks in various
fields have some common topology characteristics. They have a small
average distance like random graphs, a large clustering coefficient
and power-law degree distribution \cite{WS98,BA99}, which are called
the small-world and scale-free characteristics. The other
characteristic is that the social networks are assortative while
almost all biological and technological networks are opposite. The
networks with high clustering and small average distance are the
small-world model of Watts and Strogatz (WS)\cite{WS98}, while the
networks with power-law degree distribution are the scale-free
network model of Barab$\acute{a}$si and Albert (BA) \cite{BA99}. The
BA model is a pioneering work in the studies on networks, which
suggests that the growth and preferential attachment are two main
self-organization mechanisms. Although BA model can generate the
power-law degree distributions, its assortative coefficient $r$
equals to zero in the limit of large size thus fail to reproduce the
disassortative property that extensively exists in the real-world
networks. Recently, some models that can generate either assortative
or disassortative networks have been reported
\cite{17,18,19,15,WWX2,WWX}. Wang {\it et al.} presented a mutual
attraction model for both assortative and disassortative weighted
networks. The model found that the initial attraction $A$ of the
newly added nodes may contribute to the difference of the
assortative and disassortative networks \cite{WWX2}. Liu {\it et
al.} \cite{cpl} proposed a self-learning mutual selection model for
weighted networks, which demonstrated that the self-learning
probability $p$ may be the reason why the social networks are
assortative and the technological networks are disassortative.
However, one should not expect the existence of a omnipotent model
that can completely illuminate the underlying mechanisms for the
emergence of disassortative property in various network systems. In
this paper, beside the previous studies, we exhibit an alternative
model that can generate scale-free small-world networks with tunable
assortative coefficient, which may shed some light in finding the
possible explanations to the different evolution mechanisms between
assortative and disassortative networks.

Dorogovtsev {\it et. al} \cite{M6} proposed a simple model of
scale-free growing networks . In this model, a new node is added to
the network at each time step, which connects to both ends of a
randomly chosen link undirected. The model can be equally described
by the process that the newly added node connect to node $i$
preferentially, then select a neighbor node of the node $i$
randomly. Holme and Kim \cite{M7} proposed a model to generate
growing scale-free networks with tunable clustering. The model
introduced an additional step to get the trial information and
demonstrated that the average number of trial information controls
the clustering coefficient of the network. It should be noticed that
the newly added node connect to the first node $i$ preferentially,
while connect to the neighbor node of the first node $i$ randomly.
In this paper, we will propose a growing scale-free network model
with tunable assortative coefficient. Inspired by the above two
models, the new node is added into the network by two steps. In the
first step, the newly added node connects to the existing nodes $i$
preferentially. In the second step, this node selects a neighbor
node $s$ of the node $i$ with probability $k_s^{\beta}/\sum_{j\in
{\Gamma_i}} k_j^{\beta}$, where $\beta$ is the parameter named
preferential exponent and $\Gamma_i$ is the neighbor node set of
node $i$. This model will be equal to the Holme-Kim model[37] when
$\beta=0$, and the MRGN model[35] when $\beta=1$  Specifically, the
model can generate a nontrivial clustering property and tunable
assortativity coefficient. Therefore, one may find explanations to
various real-world networks by our microscopic mechanisms.

\section{Construction of the Model}
\medskip

Our model is defined as following.

\begin{description}
\item[(1)] Initial condition: The model starts with $m_{0}$
connected nodes.\\[2pt]

\item[(2)] Growth: At each time step, one new node $v$ with $m$
edges is added at every time step. Time  $t$ is identified as the
number of time steps.

\item[(3)] The first step: Each edge of $v$ is then
attached to an existing node with the probability proportional to
its degree, i.e., the probability for a node $i$ to be attached to
$v$ is
\begin{equation}
\Pi_i=\frac{k_i}{\sum_j k_j}.
\end{equation}

\item[(4)] The second step: If an
edge between $v$ and $w$ was added in the first step, then add one
more edge from $v$ to a randomly chosen neighbor $s$ of $w$ with
probability $P_t$ according to the following probability
\begin{equation}
p_{s}=\frac{k_{s}^{\beta}}{\sum_{v\in \Gamma_i}{k_{v}^{\beta}}},
\end{equation}
If there remains no pair to connect, i.e., if all neighbors of $w$
were always connected to $v$, do the first step instead.

 \end{description}

\section{Characteristics of the Model}
\medskip

\subsection{Degree distribution}
The degree distribution is one of the most important statistical
characteristics of networks. Since some real-world networks are
scale-free, whether the network is of the power-law degree
distribution is a criterion to judge the validity of the model. By
adopting the mean-field theory, the degree evolution of individual
node can be described as
\begin{equation}\label{F2.1}
\frac{\partial k_i}{\partial t}=P(i)+\sum_{j\in
\Gamma_i}P(i|j)P(j),
\end{equation}
where $P(i)$ denotes the probability that the node $i$ with degree
$k_i$ is selected at the first step, $P(i|j)$ denotes the
conditional probability that node $i$ is a neighbor of node $j$ with
degree $k_j$ which has been selected at the first step.

According to the preferential attachment mechanism of the first
step, one has
\begin{equation}\label{F2.2}
P(i)=\frac{k_i}{\sum_j k_j}.
\end{equation}
The conditional probability $P(i|j)$ can be calculated by
\begin{equation}\label{F2.3a}
P(i|j)=\frac{k_i}{\sum_{l\in \Gamma_j}k_l}.
\end{equation}
According to the second step, one has that
\begin{equation}\label{F2.5}
\frac{\partial k_i}{\partial t}=\frac{k_i}{\sum_l k_l}+\sum_{j\in
\Gamma_i}\Big(\frac{k_i^{\beta}}{\sum_{s\in \Gamma_j}
k_s^{\beta}}\frac{k_j}{\sum_l k_l}\Big).
\end{equation}
If $\beta=0$, we get that
\begin{equation}\label{F2.6}
\frac{\partial k_i}{\partial t}=\frac{k_i}{\sum_l k_l}+\sum_{j\in
\Gamma_i}\Big(\frac{1}{k_j}\frac{k_j}{\sum_l
k_l}\Big)=\frac{2k_i}{\sum_l k_l}.
\end{equation}
Then we can get that $P(k)\sim k^{-3}$, which has been proved by
Holme and Kim \cite{M7}. If $\beta=1$, the following formula can be
obtained under the assumption that the present network is
non-assortative.
\begin{equation}\label{F2.7} \frac{\partial
k_i}{\partial t}=\frac{k_i}{\sum_j k_j}+\sum_{j\in
\Gamma_i}\frac{k_i}{\langle k\rangle k_l}\frac{k_l}{\sum_j
k_j}=\frac{2k_i}{\sum_j k_j}.
\end{equation}
We can obtain that the degree distribution $p(k)\sim k^{-r}$ obeys
the power-law and the exponent $\gamma=3$. The numerical results are
demonstrated in Fig. 1. From Fig. 1, we can get that the exponents
$\gamma$ of the degree distribution are around -3 when
$\beta=-2,-1,0,1,2$. When $\beta>0$, the exponent $\gamma$ would
increase slightly as the $\beta$ increases.

\begin{figure}[ht]
  \begin{center}
       \center \includegraphics[width=13cm]{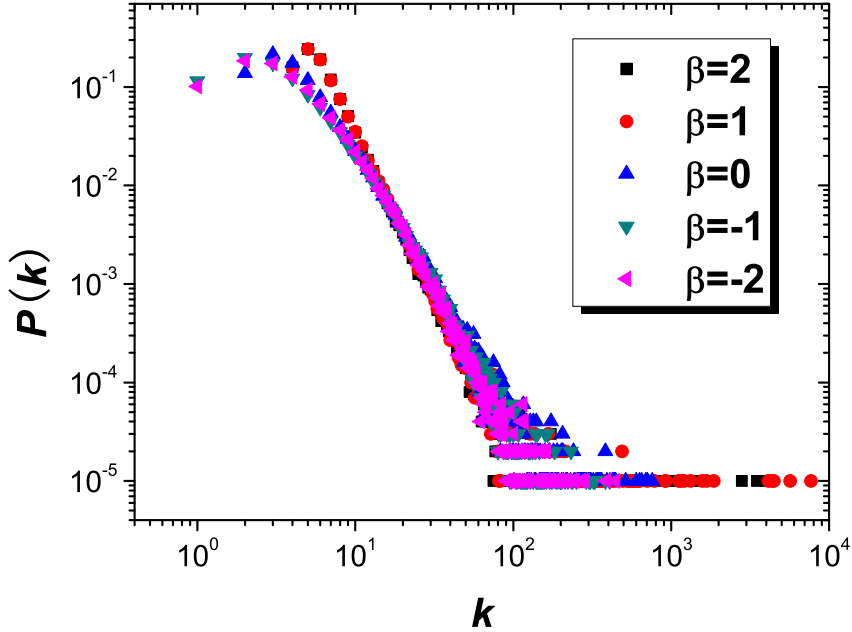}
       \caption{(Color Online) Degree distribution of the present network with $N=100000$ nodes  when $m=3$ and $P_t=0.3$.
       In this figure, $p(k)$ denotes the probability of nodes with
       degree $k$ in the network. The power-law degree distribution exponents $\gamma$ of the
       four probability density function are $\gamma_{\beta=2}=3.11\pm 0.05$,
       $\gamma_{\beta=1}=3.11\pm 0.05$, $\gamma_{\beta=0}=2.93\pm 0.04$, $\gamma_{\beta=-1}=2.96\pm 0.05$ and
        $\gamma_{\beta=-2}=2.95\pm 0.04$.}
 \end{center}
\end{figure}

\begin{figure}
  \begin{center}
       \center \includegraphics[width=13cm]{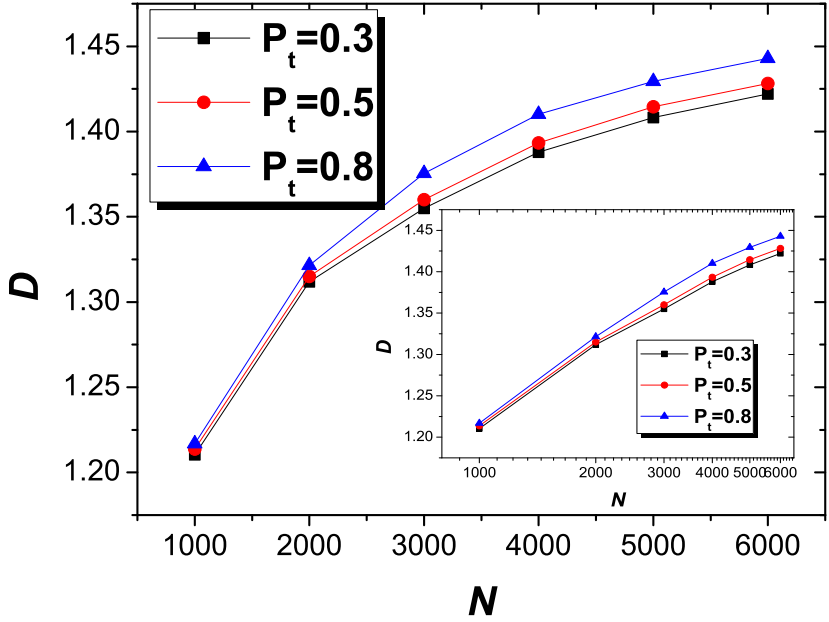}
       \caption{(Color Online) The dependence between the average distance $D$ and
the order $N$ of the present network, when $\beta=1$ and $m=4$. One
can see that $L$ increases very slowly as ${\rm ln}N$ increases. The
inset exhibits the curve where $L$ is considered as a function of
$N$. All the data are obtained by 10 independent simulations.}
 \end{center}
\end{figure}

\subsection{Average distance}

The average distance is also one of the most important parameters
to measure the efficiency of communication networks, which is
defined as the mean distance over all pairs of nodes.  The average
distance plays a significant role in measuring the transmission
delay. Firstly, we give the following lemma \cite{M1}.

{\bf Lemma 1} {\it For any two nodes $i$ and $j$, each shortest
path from $i$ to $j$ does not pass through any nodes $k$
satisfying that $k>{\rm max}\{i,j\}$.
}

{\it Proof.} Denote the shortest path from the node $i$ to $j$ of
length $n+1$ by $i\rightarrow x^1 \rightarrow x^2 \cdots \rightarrow
x^n \rightarrow j $($SP_{ij}$),  where $n>0$. Suppose that $x^k={\rm
max}\{x^1, x^2, \cdots, x^n\}$, if $x^k\leq {\rm max}\{i, j\}$, then
the conclusion is true. If $x^k>{\rm max}\{i,j\}$, denote the
youngest node of $SP_{ij}$ by $k$. Denote the subpath passing
through node $k$ by $l\rightarrow k\rightarrow r$, where the node
$l$ and $r$ are the neighbors of node $k$, then we can prove that
node $l$ and $r$ are connected. The shortest path $SP_{ij}$ passes
from the node $l$ to $r$ directly, which is conflicted with the
hypothesis.

Let $d(i,j)$ represent the distance between node $i$ and $j$ and
$\sigma(N)$ as the total distance, i.e., $\sigma(N)=\sum_{1\leq
i<j\leq N}d(i,j)$. The average distance of the present model with
order $N$, denoted by $L(N)$, is defined as following
\begin{equation}\label{F3.1}
L(N)=\frac{2\sigma(N)}{N(N-1)}.
\end{equation}
According to Lemma 1, the newly added node will not affect the
distance between the existing ones. Hence we have
\begin{equation}\label{F3.2}
\sigma(N+1)=\sigma(N)+\sum^N_{i=1}d(i,N+1).
\end{equation}
Assume that the $(N+1)$th node is added to the edge $E_{y_1y_2}$,
then Equ. (\ref{F3.2}) can be rewritten as
\begin{equation}\label{F3.3}
\sigma(N+1)=\sigma(N)+N+\sum^N_{i=1}D(i,y),
\end{equation}
where $D(i,y)={\rm min}\{d(i,y_1),d(i,y_2)\}$. Denote $y$ as the
edge connected the node $y_1$ and $y_2$ continuously, then we have
the following equation
\begin{equation}\label{F3.4}
\sigma(N+1)=\sigma(N)+N+\sum_{i=\Lambda}d(i,y),
\end{equation}
where the node set $\Lambda=\{1,2, \cdots, N\}-\{y_1, y_2\}$ has
$(N-2)$ members. The sum $\sum_{i=\Lambda}d(i,y)$ can be considered
as the distance from each node of the network to node $y$ in the
present model with order $N-2$. Approximately, the sum
$\sum_{i=\Lambda}d(i,y)$ is equal to $L(N-2)$. Hence we have
\begin{equation}\label{F3.5}
\sum_{i=\Lambda}d(i,y)\approx (N-2)L(N-2).
\end{equation}
Because the average distance $L(N)$ increases monotonously with
$N$, this yields
\begin{equation}\label{F3.6}
(N-2)L(N-2)=(N-2)\frac{2\sigma(N-2)}{(N-2)(N-3)}<\frac{2\sigma(N)}{N-1}.
\end{equation}
Then we can obtain the inequality
\begin{equation}\label{F3.7}
\sigma(N+1)<\sigma(N)+N+\frac{2\sigma(N)}{N-1}.
\end{equation}
Enlarge $\sigma(N)$, then the upper bound of the increasing
tendency of $\sigma(N)$ reads
\begin{equation}\label{F3.8}
\frac{d\sigma(N)}{dN}=N+\frac{2\sigma(N)}{N-1}.
\end{equation}
This leads to the following solution
\begin{equation}\label{F3.9}
\sigma(N)=(N-1)^2{\rm log}(N-1)+C_1(N-1)^2-(N-1).
\end{equation}
The numerical results are demonstrated in Fig. 2.

\subsection{Clustering property}

\begin{figure}
  \begin{center}
       \center \includegraphics[width=12cm]{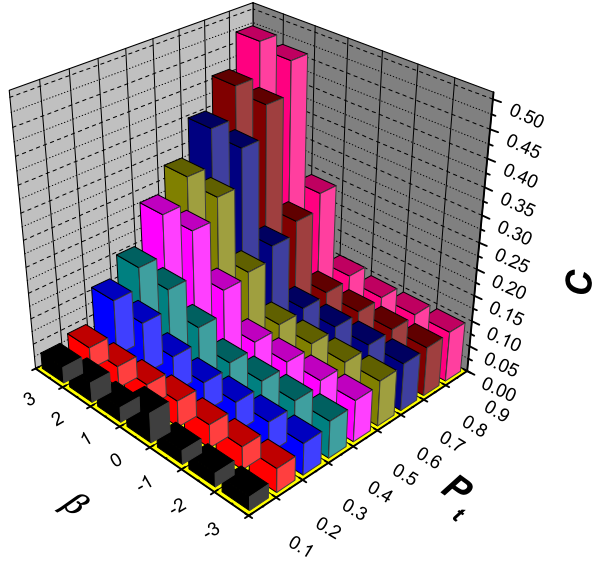}
       \caption{(Color Online) The scale of $C$ with various $\beta$ and $P_t$ when $m=3$. The data are averaged over 10 independent runs of network size of $N=6000$.}
 \end{center}
\end{figure}
The small-world characteristic consists of two properties: large
clustering coefficient and small average distance. The clustering
coefficient, denoted by $C$, is defined as
$C=\sum_{i=1}^N\frac{C_i}{N}$, where $C_i$ is the local clustering
coefficient for node $i$. $C_i$ is
\begin{equation}\label{F4.1}
C_i=\frac{2E(i)}{k_i(k_i-1)},
\end{equation}
where $E(i)$ is the number of edges in the neighbor set of the
node $i$, and $k_i$ is the degree of node $i$. When the node $i$
is added to the network, it is of degree $m+mP_t$ and $E(i)=mP_t$.
If a new node is added to be a neighbor of $i$ at some time step,
$E(i)$ will increase by $mP_t$ since the newly added node will
connect with one of the neighbors of the node $i$ with probability
$P_t$. Therefore, in terms of $k_i$, the expression of $E(i)$ can
be written as
\begin{equation}\label{F4.2}
E(i)=mP_t+P_t[k_i-(m+mP_t)].
\end{equation}
Hence, we have that
\begin{equation}\label{F4.3}
C_i=\frac{2[mP_t+P_t(k_i-m-mP_t)]}{k_i(k_i-1)}=2\Big(\frac{mP_t^2}{k_i}+\frac{P_t-mP_t^2}{k_i-1}\Big).
\end{equation}
This expression indicates that the local clustering scales as
$C(k)\sim k^{-1}$, where $C(k)$ denotes the average clustering
coefficient value of nodes with degree $k$. It is interesting that a
similar scaling has been observed in many artificial models
\cite{M7,M1,M2,M8} and several real-world networks \cite{M9}. The
degree-dependent average clustering coefficient $C(k)$ has been
demonstrated in Fig. 4. Consequently, we have
\begin{equation}\label{F4.4}
C=\frac{2}{N}\sum^N_{i=1}\Big(\frac{mP_t^2}{k_i}+\frac{P_t-mP_t^2}{k_i-1}\Big).
\end{equation}
\begin{figure}
  \begin{center}
       \center \includegraphics[width=12cm]{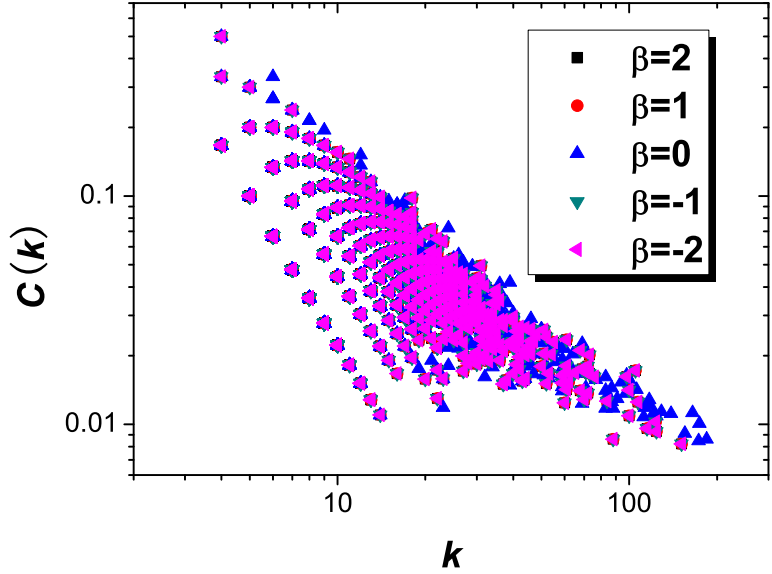}
       \caption{(Color Online) $C(k)$ vs $k$ to various $\beta$ and when $m=3$ and $P_t=0.3$.
        The data are averaged over 10 independent runs of network size of $N=6000$.}
 \end{center}
\end{figure}
Since the degree distribution is $P(k)=a k^{-3}$, where $k=k_{\rm
min},\cdots, k_{\rm max}$. The constant $a$ satisfies the
normalization equation
\begin{equation}
\sum_{k_{\rm min}}^{k_{\rm max}}ak^{-3}=1,
\end{equation}
one can get that $a=2k_{\rm min}^2$. The average clustering
coefficient $C$ can be rewritten as
\begin{equation}
\begin{array}{rcl}
C & = & \frac{2}{N}\sum_{k_{\rm min}}^{k_{\rm max
}}\Big(\frac{NP(k)mP_t^2}{k}+\frac{NP(k)(P_t-mP_t^2)}{k-1}\Big)\\[2pt]
          & \approx & 2\sum_{k_{\rm min}}^{k_{\rm max
}}\Big(ak^{-4}mP_t^2+ak^{-3}(P_t-mP_t^2)/(k-1)\Big)\\[2pt]
\end{array}
\end{equation}
The numerical results are demonstrated in Fig.3. From figure 3, we
can get that if $\beta\leq 0$, the numerical results are consistent
with the theoretical predictions approximately, while if $\beta>0$,
the fluctuations emerges. The departure from analysis results is
observed, which may attribute to the fluctuations of the power-law
exponent of degree distribution. It is also helpful to compare the
present method with previous analysis approaches on clustering
coefficient for Holme-Kim model \cite{AD1,AD2}.

\subsection{Assortative coefficient}
The assortative coefficient $r$ can be calculated from
\begin{equation}
r=\frac{M^{-1}\sum_i j_i
k_i-[M^{-1}\sum_i\frac{1}{2}(j_i+k_i)]^2}{M^{-1}\sum_i\frac{1}{2}(j_i^2+k_i^2)-[M^{-1}\sum_i\frac{1}{2}(j_i+k_i)]^2},
\end{equation}
where $j_i$, $k_i$ are the degrees of the vertices of the $i$th
edge, for  $i=1, 2, \cdots,M$ \cite{r,r2}.

\begin{figure}
  \begin{center}
       \center \includegraphics[width=12cm]{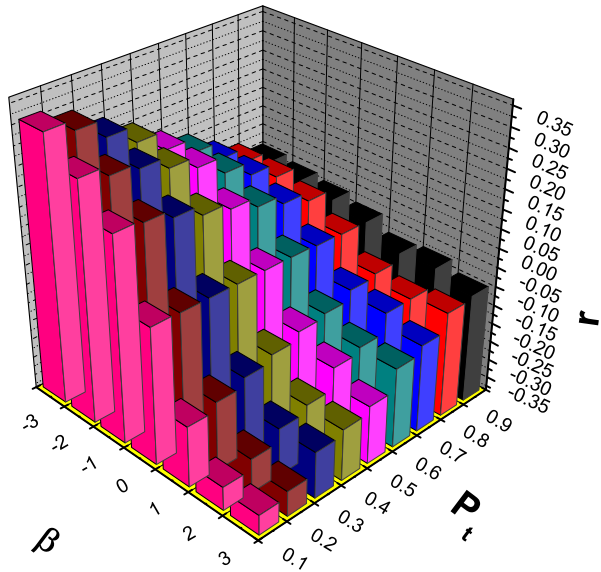}
       \caption{(Color Online) The scale of $r$ with various $\beta$ and $P_t$ when $m=1$.
       The data are averaged over 10 independent runs of network size of $N=6000$.}
 \end{center}
\end{figure}

From Fig. 5, we can find that when $\beta>0$ the assortative
coefficient $r$ increases with the probability $P_t$, while $r$
decreases with the probability $P_t$ when $\beta<0$. As $\beta=0$,
$r$ equals to zero approximately.

\section{Conclusion and Discussion}
In this paper, we propose a simple rule that generates scale-free
small-world networks with tunable assortative coefficient. The
inspiration of this model is to introduce the parameter $\beta$ to
Holme-Kim model. The simulation results are consistent with the
theoretical predictions approximately. Interestingly, we obtain the
nontrivial clustering coefficient $C$ and tunable degree
assortativity $r$, depending on the parameters $\beta$. The model
can unify the characterization of both assortative and
disassortative networks. Specially, studying the degree-dependent
average clustering coefficient $C(k)$ also provides us with a better
description of the hierarchies and organizational architecture of
weighted networks. Our model may be conducive to future
understanding or characterizing real-world networks.

This work has been supported by the Chinese Natural Science
Foundation of China under Grant Nos. 70431001, 70271046 and
70471033.

\end{document}